\documentclass[pdflatex,sn-mathphys-num]{sn-jnl}

\usepackage{graphicx}%
\usepackage{multirow}%
\usepackage{amsmath,amssymb,amsfonts}%
\usepackage{amsthm}%
\usepackage{mathrsfs}%
\usepackage[title]{appendix}%
\usepackage{textcomp}%
\usepackage{booktabs}%
\usepackage{braket}%

\theoremstyle{thmstyleone}%
\newtheorem{theorem}{Theorem}%

\theoremstyle{thmstyletwo}%

\theoremstyle{thmstylethree}%

\begin{document}

\title[Article Title]{A simple model of quantum walk with a gap in distribution}

\author*[1]{\fnm{Takuya} \sur{Machida}}\email{machida.takuya@nihon-u.ac.jp}

\affil*[1]{\orgdiv{College of Industrial Technology}, \orgname{Nihon University}, \orgaddress{\city{Narashino}, \postcode{275-8576}, \state{Chiba}, \country{Japan}}}

\abstract{%
Quantum walks are quantum counterparts of random walks and their probability distributions are different from each other.
A quantum walker distributes on a Hilbert space and it is observed at a location with a probability.
The finding probabilities have been investigated and some interesting things have been analytically discovered.
They are, for instance, ballistic behavior, localization, or a gap.
We study a 1-dimensional quantum walk in this paper.
Although the walker launches off at a location under a localized initial state, some numerical experiments show that the quantum walker does not seem to distribute around the launching location, which suggests that the probability distribution holds a gap around the launching location.
To prove the gap analytically, we derive a long-time limit distribution, from which one can tell more details about the finding probability.
}


\maketitle

\section{Introduction}
Quantum walks are quantum analogies of random walks in mathematics.
The quantum walkers have some inner states which are interpreted as spin states in physics, and move in space in superposition.
Starting with an initial state, a walker repeats updating its system with unitary operations and is observed at a location with a finding probability.
Although quantum walks were introduced in mathematics, physics, or quantum computation before 2000~\cite{Gudder1988,AharonovDavidovichZagury1993,Meyer1996}, they began to get attention in science around 2000 due to the studies of quantum computers.
As well known, Grover's algorithm can find targets much faster than the corresponding classical algorithm, and it can be considered as a quantum walk on a graph~\cite{Grover1996}.
For database search in quantum computers, quantum walks have been focused on in quantum informations~\cite{Venegas-Andraca2008}.

The major study in quantum walks is their finding probabilities and they have been investigated numerically or analytically.
Analytic researches have resulted in long-time limit theorems in mathematics.
The limit theorems are useful to understand how the quantum walkers distribute in space after they have repeated their unitary processes a lot of times.
The first limit theorem was proved in 2002 and it stated a long-time limit distribution which was different from the limit distributions of classical random walks~\cite{Konno2002a}.
Since then, many limit theorems have been demonstrated and they told us interesting behavior of quantum walks~\cite{Venegas-Andraca2012}.

In 2015, a remarkable behavior was discovered~\cite{GrunbaumMachida2015}.
A quantum walk held a gap in distribution and the walker was never observed around the launching location.
After the paper had been published, two types of quantum walk were reported and they also had a gap in distribution.
One of them was a 3-state quantum walk in 2006 and its probability distribution showed a gap and localization~\cite{Machida2016b}.
The other was a 2-state walk in 2018 and its model was complicated~\cite{Machida2018}.
Motivated by the quantum walk in 2018, we try to seek a simpler model which makes a probability distribution with a gap.
We define a quantum walk in the next section and numerically observe its probability distribution.
Since the numerical experiments let us expect that the quantum walk distributes with a gap, we will work on limit distributions in Sect.~\ref{sec:limit_theorem} so that the quantum walk is proved to hold a gap in distribution.

\section{Definition of a quantum walk}
\label{sec:introduction}
\begin{figure}[h]
\begin{center}
 \begin{minipage}{100mm}
  \begin{center}
   \includegraphics[scale=0.5]{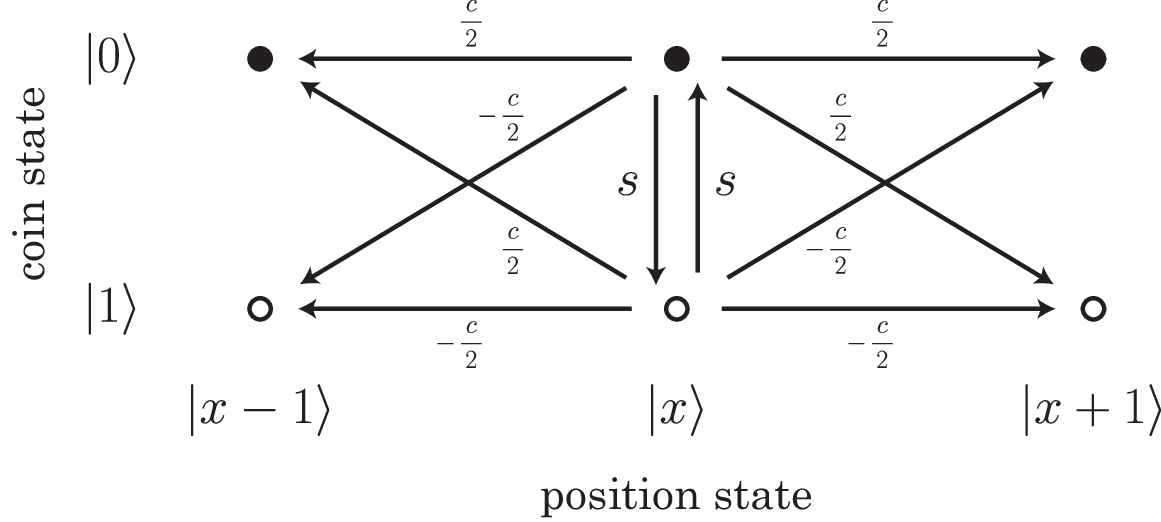}\\[2mm]
  (a) unitary operation $U_1$
  \end{center}
 \end{minipage}
 \bigskip\bigskip

 \begin{minipage}{100mm}
  \begin{center}
   \includegraphics[scale=0.5]{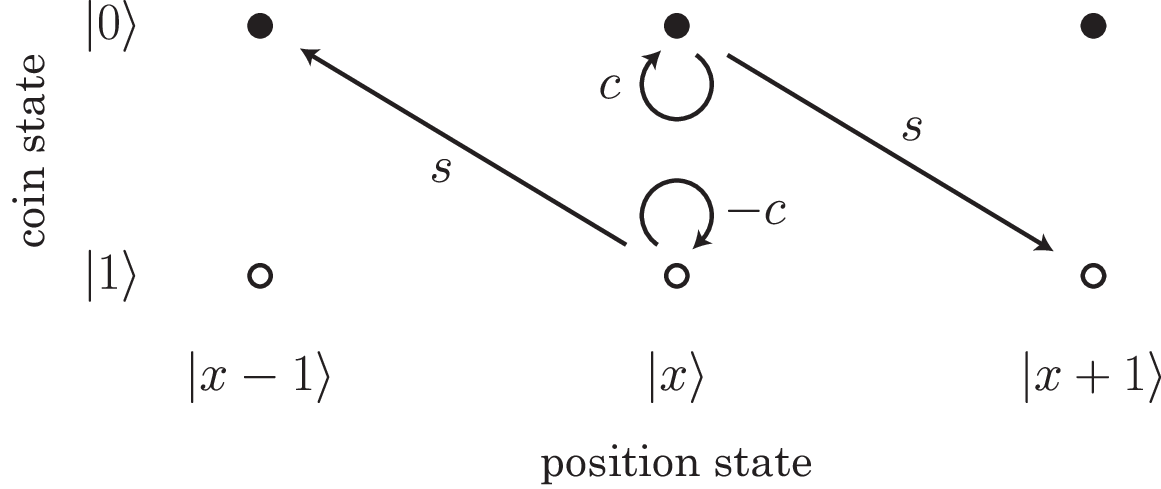}\\[2mm]
  (b) unitary operation $U_2$
  \end{center}
 \end{minipage}
 \bigskip

\caption{The walker at position $x\in\mathbb{Z}$ shifts to other positions and stays at the same position, changing its coin states $\ket{0}$ and $\ket{1}$. The mathematical description is explained in Eq.~\eqref{eq:200807-1} as unitary operations $U_1$ and $U_2$.}
\label{fig:1}
\end{center}
\end{figure}

The quantum walker with two coin states $\ket{0}$ and $\ket{1}$ is supposed to locate at points, whose set is represented by $\mathbb{Z}=\left\{0,\pm 1,\pm 2,\ldots\right\}$, in superposition.
Its system is described on a tensor Hilbert space $\mathcal{H}_p\otimes\mathcal{H}_c$.
The Hilbert space $\mathcal{H}_p$ represents the locations and it is spanned by the orthogonal normalized basis $\left\{\ket{x} : x\in\mathbb{Z}\right\}$.
Also, the Hilbert space $\mathcal{H}_c$ represents the coin states and it is spanned by the orthogonal normalized basis $\left\{\ket{0},\ket{1}\right\}$.
Let us define
\begin{equation}
 \ket{0}=\begin{bmatrix}
	  1\\0
	 \end{bmatrix},\quad
	 \ket{1}=\begin{bmatrix}
		  0\\1
		 \end{bmatrix},
\end{equation}
for the Hilbert space $\mathcal{H}_c$ in this paper.
Given two unitary operations, the coin states and the position states of the walker are operated with them.
More precisely, the system of quantum walk at time $t\,(=0,1,2,\ldots)$, represented by $\ket{\Psi_t}\in\mathcal{H}_p\otimes\mathcal{H}_c$, gets a new system at time $t+1$, represeted by $\ket{\Psi_{t+1}}\in\mathcal{H}_p\otimes\mathcal{H}_c$, with unitary operations $U_1$ and $U_2$,  
\begin{equation}
 \ket{\Psi_{t+1}}=U_2U_1\ket{\Psi_t},\label{eq:200807-1}
\end{equation}
\clearpage
where
\begin{align}
 U_1=&\sum_{x\in\mathbb{Z}}\ket{x-1}\bra{x}\otimes\frac{\cos\theta}{2}\,\Bigl(\ket{0}\bra{0}-\ket{1}\bra{0}+\ket{0}\bra{1}-\ket{1}\bra{1}\Bigr)\nonumber\\
 &+\ket{x}\bra{x}\otimes\sin\theta\,\Bigl(\ket{0}\bra{1}+\ket{1}\bra{0}\Bigr)\nonumber\\
 &+\ket{x+1}\bra{x}\otimes\frac{\cos\theta}{2}\,\Bigl(\ket{0}\bra{0}+\ket{1}\bra{0}-\ket{0}\bra{1}-\ket{1}\bra{1}\Bigr),\\[3mm]
 U_2=&\sum_{x\in\mathbb{Z}}\ket{x-1}\bra{x}\otimes\sin\theta\,\ket{0}\bra{1}+\ket{x}\bra{x}\otimes\cos\theta\Bigl(\ket{0}\bra{0}-\ket{1}\bra{1}\Bigr)\nonumber\\
 &+\ket{x+1}\bra{x}\otimes\sin\theta\,\ket{1}\bra{0}.
\end{align}
The parameter $\theta$ characterizes the unitary operations and it is fixed at a value on the interval $(0,\pi/2)\cup (\pi/2, \pi)$.
Figure~\ref{fig:1} depicts how the coin states and the position states are changed by these operations.
In the picture, the notations $c$ and $s$ are short for $\cos\theta$ and $\sin\theta$, respectively.
Note that the product of two unitary operations $U_1$ and $U_2$, that is $U_2U_1$, updates the system of quantum walk at each time. 

\begin{figure}[h]
\begin{center}
 \begin{minipage}{60mm}
  \begin{center}
   \includegraphics[scale=0.3]{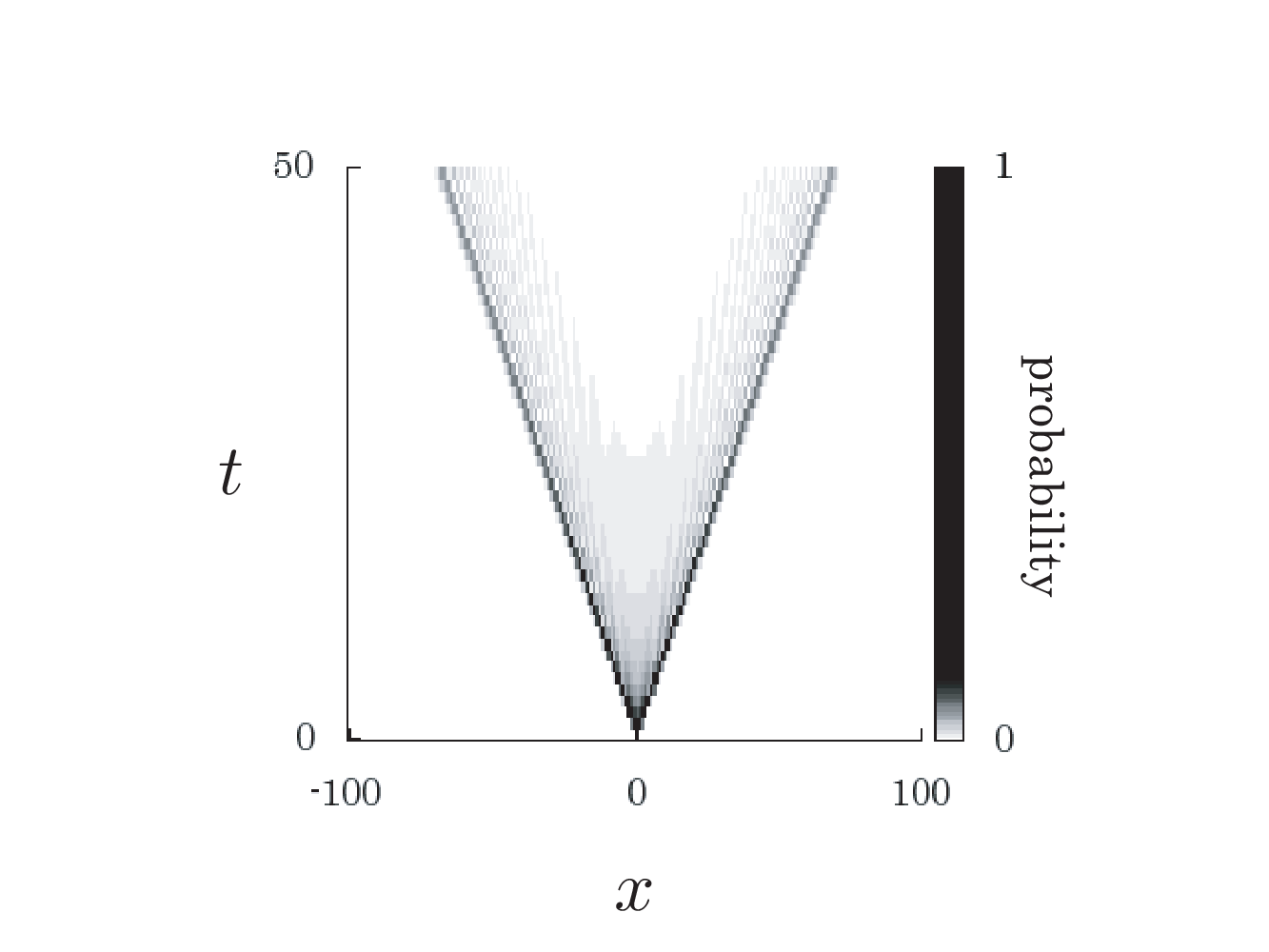}\\[2mm]
  (a) $\theta=\pi/4$
  \end{center}
 \end{minipage}
 \begin{minipage}{60mm}
  \begin{center}
   \includegraphics[scale=0.3]{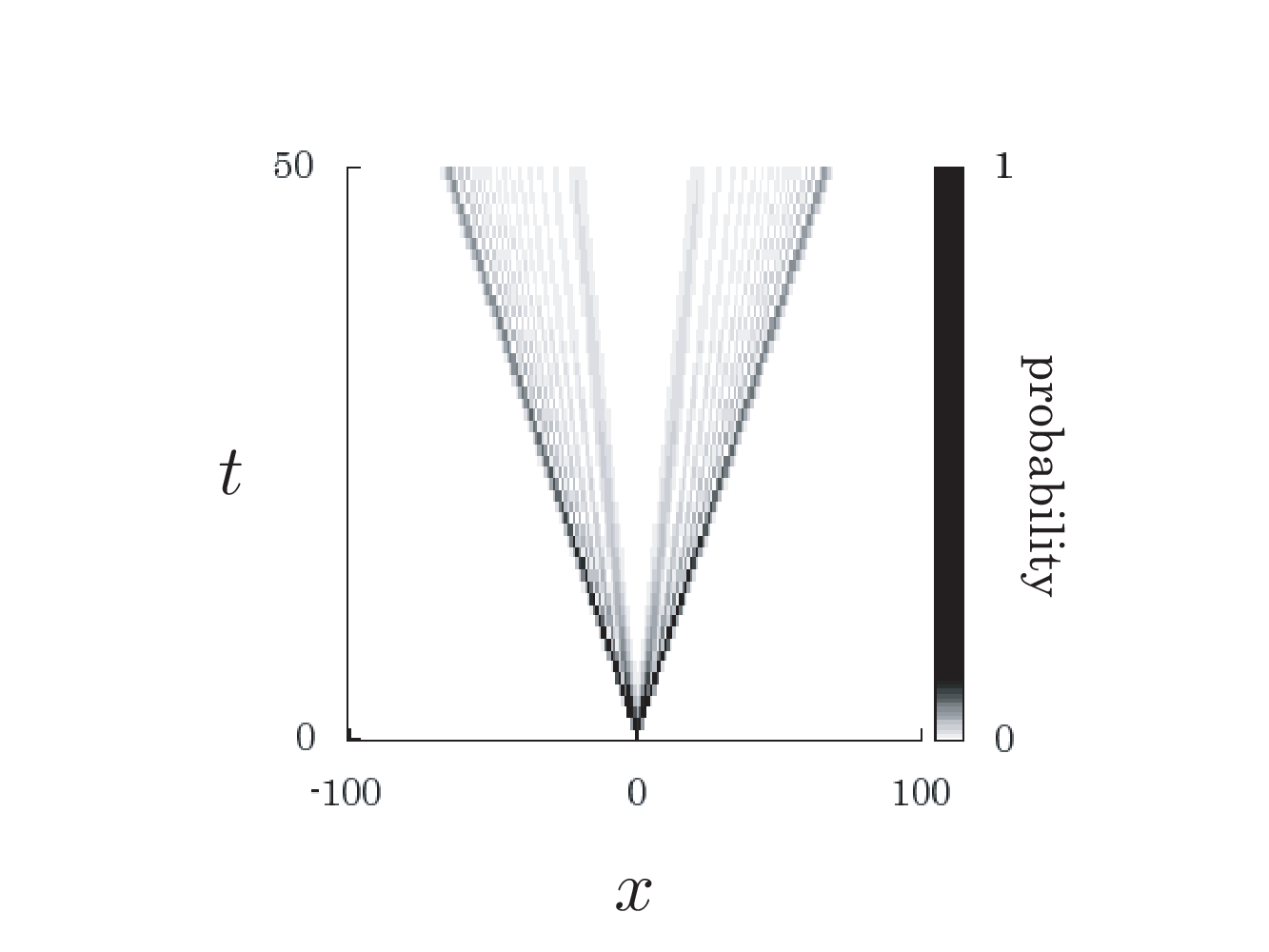}\\[2mm]
  (b) $\theta=\pi/6$
  \end{center}
 \end{minipage}
 \bigskip

\caption{The probability distribution $\mathbb{P}(X_t=x)$ can be splitting to two major parts as the walker is getting updated. The walker launches off with the localized initial state at the origin, $\ket{\Psi_0}=\ket{0}\otimes (1/\sqrt{2}\ket{0}+i/\sqrt{2}\ket{1})$.}
\label{fig:2}
\end{center}
\end{figure}

\begin{figure}[h]
\begin{center}
 \begin{minipage}{40mm}
  \begin{center}
   \includegraphics[scale=0.2]{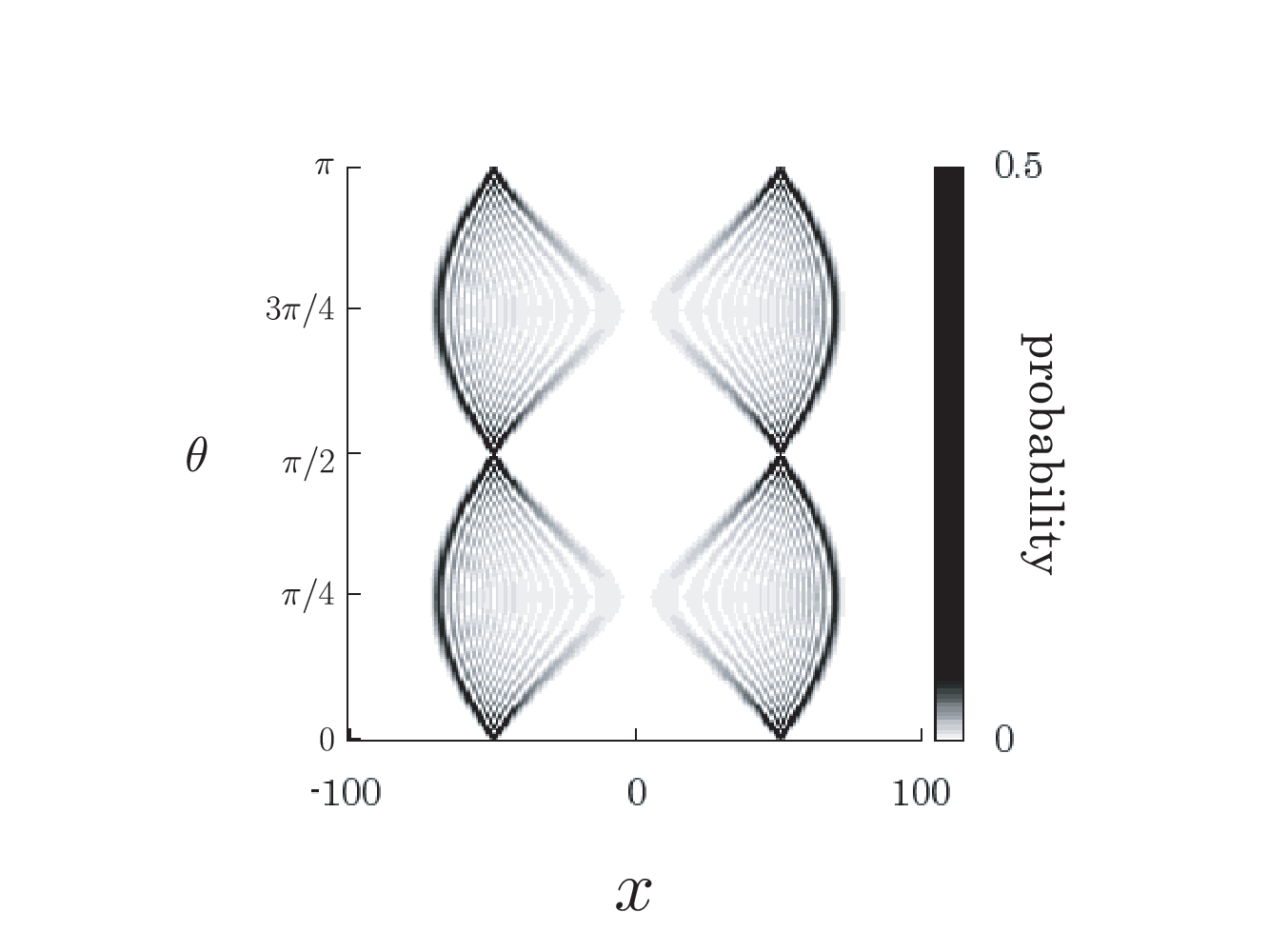}\\[2mm]
  (a) $\alpha=1/\sqrt{2},\, \beta=i/\sqrt{2}$
  \end{center}
 \end{minipage}
 \begin{minipage}{40mm}
  \begin{center}
   \includegraphics[scale=0.2]{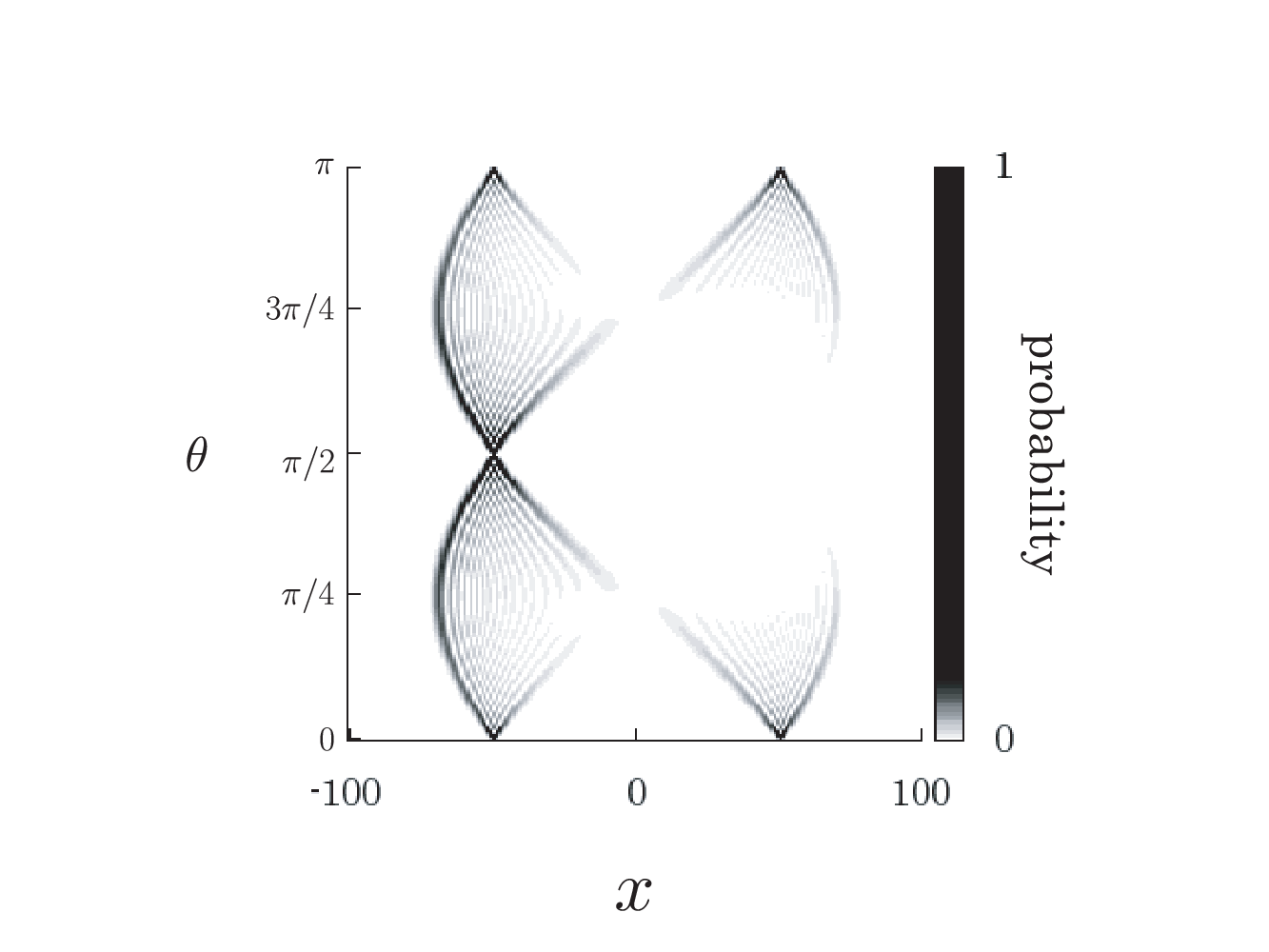}\\[2mm]
  (b) $\alpha=1,\, \beta=0$
  \end{center}
 \end{minipage}
 \begin{minipage}{40mm}
  \begin{center}
   \includegraphics[scale=0.2]{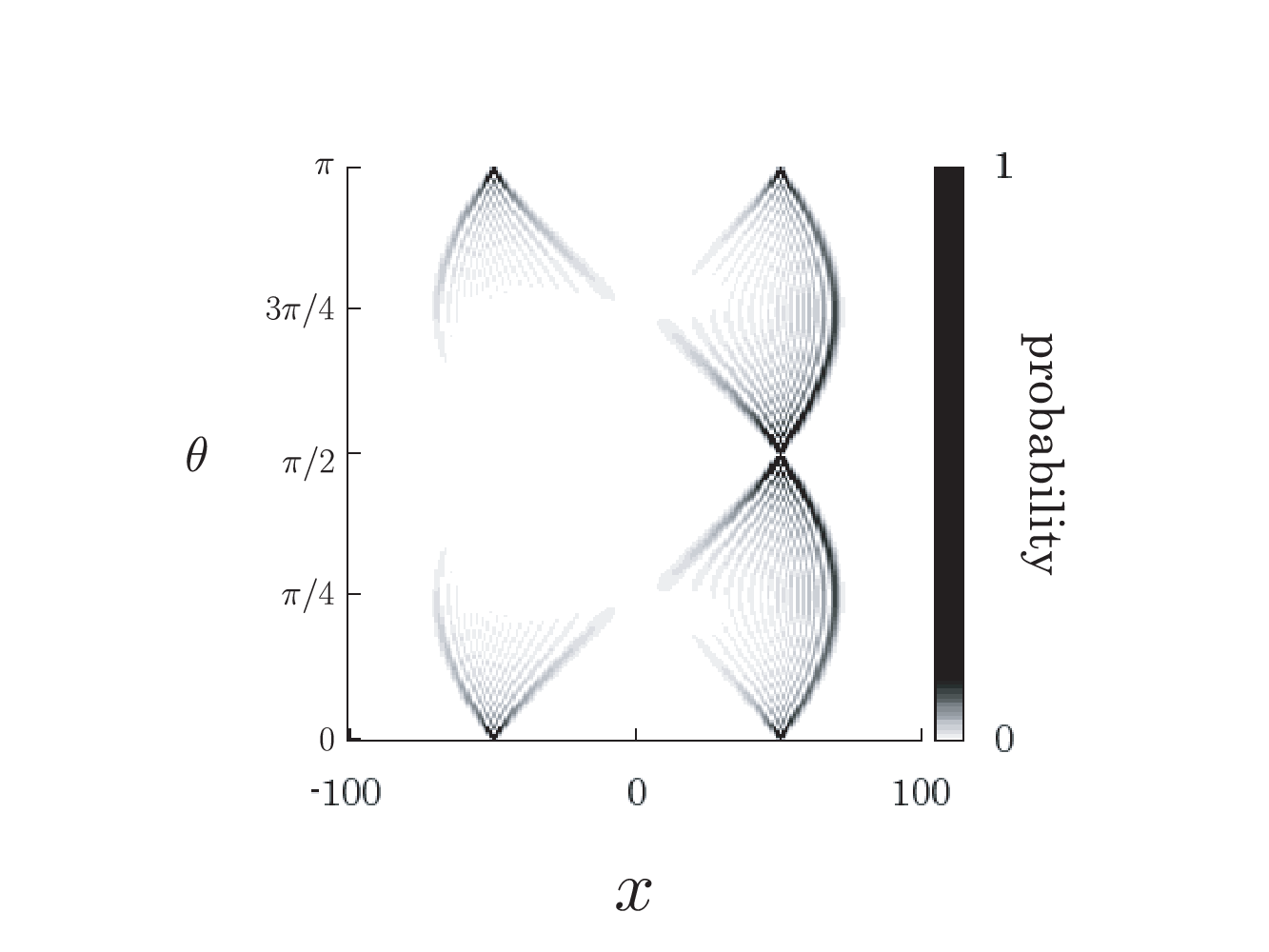}\\[2mm]
  (c) $\alpha=0,\, \beta=1$
  \end{center}
 \end{minipage}
 \bigskip

\caption{The probability distribution $\mathbb{P}(X_t=x)$ at time $50$ holds a gap which appears between two major parts, and the width of the gap depends on the value of parameter $\theta$ which characterizes the unitary operations $U_1$ and $U_2$. The initial state of the walker is given as $\ket{\Psi_0}=\ket{0}\otimes (\alpha\ket{0}+\beta\ket{1})$.}
\label{fig:3}
\end{center}
\end{figure}

The walker is assumed to launch off at the location $x=0$ as a localized initial state $\ket{\Psi_0}=\ket{0}\otimes\left(\alpha\ket{0}+\beta\ket{1}\right)\,(=\ket{0}\otimes\ket{\phi})$ where the complex numbers $\alpha$ and $\beta$ should be under the constraint $|\alpha|^2+|\beta|^2=1$.
The quantum walker is observed at position $x\in\mathbb{Z}$ at time $t\in\left\{0,1,2,\ldots\right\}$ with probability
\begin{equation}
 \mathbb{P}(X_t=x)=\bra{\Psi_t}\left\{\ket{x}\bra{x}\otimes(\ket{0}\bra{0}+\ket{1}\bra{1})\right\}\ket{\Psi_t},\label{eq:finding_prob}
\end{equation}
where $X_t$ denotes the position of the walker at time $t$.
Looking at some numerical experiments for the probability distribution, we expect that the quantum walk has a gap in distribution.
Although the walker distributes without any gap in Fig.~\ref{fig:2}-(a), we easily find a gap in Fig.~\ref{fig:2}-(b).
Also, it seems that the probability distribution holds a gap unless the parameter $\theta$ is fixed at $\pi/4$ or $3\pi/4$, as shown in Fig.~\ref{fig:3}.
To check this conjecture, we will try to get a long-time limit distribution in the next section.

Here, we see the Fourier transform of the quantum walk, which will be used to compute a limit distribution as $t\to\infty$.
Let $i$ be the imaginary unit.
With the unitary operations
\begin{align}
 \hat U_1(k)=& \cos\theta\cos k\, (\ket{0}\bra{0}-\ket{1}\bra{1})\nonumber\\
 & +(\sin\theta+i\cos\theta\sin k)\,\ket{0}\bra{1}+(\sin\theta-i\cos\theta\sin k)\,\ket{1}\bra{0},
\end{align}
\begin{align}
 \hat U_2(k)=& \cos\theta\, (\ket{0}\bra{0}-\ket{1}\bra{1})+e^{ik}\sin\theta\,\ket{0}\bra{1}+e^{-ik}\sin\theta\,\ket{1}\bra{0},
\end{align}
the Fourier transform $\ket{\hat\psi_t(k)}=\sum_{x\in\mathbb{Z}}e^{-ikx}\left\{\bra{x}\otimes(\ket{0}\bra{0}+\ket{1}\bra{1})\right\}\ket{\Psi_t}\, (k\in [-\pi,\pi))$ gets a new state,
\begin{equation}
 \ket{\hat\psi_{t+1}(k)}=U_2(k)U_1(k)\ket{\hat\psi_t(k)},\label{eq:200807-2}
\end{equation}
from which
\begin{equation}
 \ket{\hat\psi_t(k)}=\left(\hat{U}_2(k)\hat{U}_1(k)\right)^t\ket{\hat\psi_0(k)}\quad (t=0,1,2,\ldots),\\
\end{equation}
follows.
The operation $\hat{U}_2(k)\hat{U}_1(k)$ is the one-step unitary operation working on the Fourier transform of the waker. 
Equation~\eqref{eq:200807-2} has come up from Eq.~\eqref{eq:200807-1}.
The initial state of the Fourier transform is computed to be $\ket{\hat\psi_0(k)}=\alpha\ket{0}+\beta\ket{1}$.
We should note that the system is reproduced by inverse Fourier transform
\begin{equation}
 \ket{\Psi_t}=\sum_{x\in\mathbb{Z}}\ket{x}\otimes\int_{-\pi}^\pi\,e^{ikx}\ket{\hat\psi_t(k)}\,\frac{dk}{2\pi}.
\end{equation}

\section{Limit theorem}
\label{sec:limit_theorem}

We will see limit theorems for the quantum walk in this section.
The statement of the theorems is expressed in two ways, depending on the value of parameter $\theta$ which is embedded in the unitary operations $U_1$ and $U_2$.

\subsection{$\theta=\pi/4, 3\pi/4$}
If we set the parameter $\theta$ at $\pi/4$ or $3\pi/4$, the quantum walk is essentially same as a quantum walk which was analyzed before.
The probability distribution $\mathbb{P}(X_t=x)$, hence, does not hold any gap in itself.

\begin{theorem}
For a real number $x$, we have
 \begin{align}
   &\lim_{t\to\infty}\mathbb{P}\left(\frac{X_t}{t}\leq x\right)\nonumber\\
   =&\int_{-\infty}^x \frac{2\sqrt{2}}{\pi(4-y^2)\sqrt{2-y^2}}\biggl\{1-\Bigl(|\alpha|^2-|\beta|^2+\alpha\overline{\beta}+\overline{\alpha}\beta\Bigr)\frac{y}{2}\biggr\}I_{(-\sqrt{2}, \sqrt{2})}(y)\,dy,
 \end{align}
 where
 \begin{align}
  I_{(-\sqrt{2}, \sqrt{2})}(x)=&\left\{\begin{array}{cl}
	   1&(x\in (-\sqrt{2}, \sqrt{2}))\\
		  0&(x\notin (-\sqrt{2}, \sqrt{2}))
		 \end{array}\right..
 \end{align}
 \label{th:limit_1}
\end{theorem}

\begin{figure}[h]
\begin{center}
 \begin{minipage}{50mm}
  \begin{center}
   \includegraphics[scale=0.4]{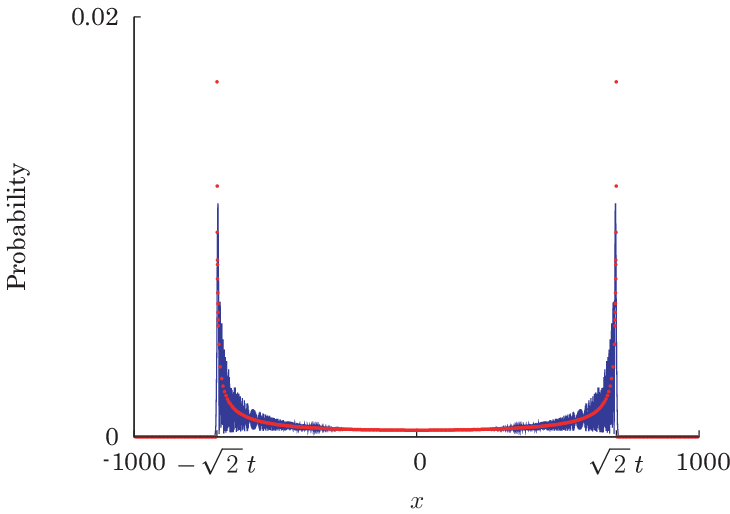}\\[2mm]
  (a) $\theta=\pi/4$
  \end{center}
 \end{minipage}
 \begin{minipage}{50mm}
  \begin{center}
   \includegraphics[scale=0.4]{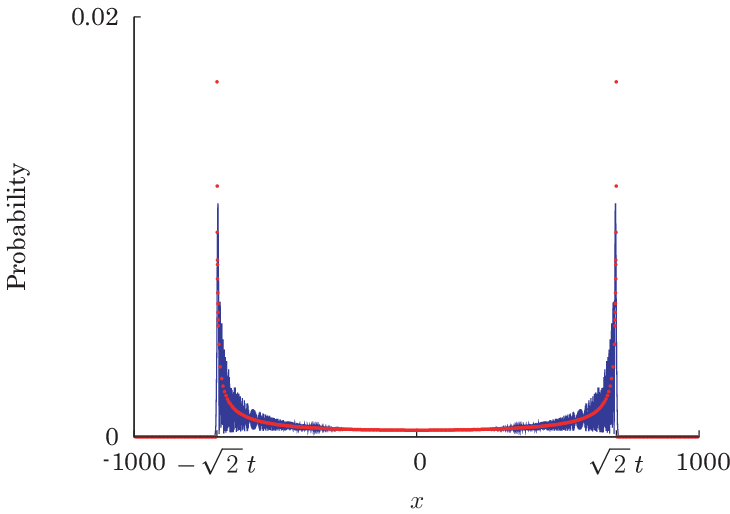}\\[2mm]
  (b) $\theta=3\pi/4$
  \end{center}
 \end{minipage}
 \bigskip

\caption{(Color figure online) The blue lines represent the probability distribution $\mathbb{P}(X_t=x)$ at time $t=500$ and the red points represent the right side of Eq.~\eqref{eq:approximation-1} as $t=500$. The limit density function approximately reproduces the probability distribution as time $t$ becomes large enough. The walker launches off at the location $x=0$ as the initial state $\ket{\Psi_0}=\ket{0}\otimes (1/\sqrt{2}\ket{0}+i/\sqrt{2}\ket{1})$.}
\label{fig:4}
\end{center}
\end{figure}

Theorem~\ref{th:limit_1} is helpful to understand the quantum walk because the limit density function
\begin{align}
 \chi_1(x)=& \frac{2\sqrt{2}}{\pi(4-x^2)\sqrt{2-x^2}}\biggl\{1-\Bigl(|\alpha|^2-|\beta|^2+\alpha\overline{\beta}+\overline{\alpha}\beta\Bigr)\frac{x}{2}\biggr\}I_{(-\sqrt{2}, \sqrt{2})}(x),
\end{align}
approximates the probability distribution $\mathbb{P}(X_t=x) \sim \frac{1}{t}\cdot \chi_1\left(x/t\right)\,(x\in\mathbb{Z})$ as time $t$ is large enough, that is more precisely,
\begin{align}
 & \mathbb{P}(X_t=x)\nonumber\\
 \sim & \frac{2\sqrt{2}\,t^2}{\pi(4t^2-x^2)\sqrt{2t^2-x^2}}\biggl\{1-\Bigl(|\alpha|^2-|\beta|^2+\alpha\overline{\beta}+\overline{\alpha}\beta\Bigr)\frac{x}{2t}\biggr\}I_{(-\sqrt{2}\,t, \sqrt{2}\,t)}(x).\label{eq:approximation-1}
\end{align}
Figure~\ref{fig:4} visualizes the approximation to the probability distribution and we numerically confirm that Theorem~\ref{th:limit_1} is appropriate.

Although a possible method to derive this limit theorem is Fourier analysis, such a way is omitted here because the computation was already demonstrated~\cite{GrimmettJansonScudo2004,Machida2016a}.
Instead, using a result in the past studies, we may explain the reason that Theorem~\ref{th:limit_1} is correct. 
When the parameter $\theta$ is fixed at $\pi/4$, the quantum walk is equivalent to a Hadamard walk because the one-step unitary operation is possible to be arranged with the $2\times 2$ Hadamard operation, represented by $H$,
\begin{equation}
 \hat{U}_2(k) \hat{U}_1(k)=\left(R\left(\frac{k}{2}\right)H\right)^4,
\end{equation}
where
\begin{align}
 R(k)=& e^{ik}\ket{0}\bra{0} + e^{-ik}\ket{1}\bra{1},\\
 H=& \frac{1}{\sqrt{2}}\Bigl(\ket{0}\bra{0} + \ket{0}\bra{1} + \ket{1}\bra{0} - \ket{1}\bra{1}\Bigr).
\end{align}
The operation $R(k)$ is comparable to an operation with which the location of the walker is shifted by $-1$ or $+1$ on the integer points $\mathbb{Z}$.
Similarly, when the parameter $\theta$ is fixed at $3\pi/4$, the one-step unitary operation is of the form
\begin{equation}
 \hat{U}_2(k) \hat{U}_1(k)=-\left(R\left(\frac{k}{2}\right)\tilde{H}\right)^4,
\end{equation}
with
\begin{equation}
 \tilde{H}=\frac{1}{\sqrt{2}}\Bigl(-\ket{0}\bra{0} - \ket{0}\bra{1} + \ket{1}\bra{0} - \ket{1}\bra{1}\Bigr).
\end{equation}
The quantum walk is, therefore, essentially same as a quantum walk whose limit theorem was already proved by some methods~\cite{Konno2002a,GrimmettJansonScudo2004,Machida2016a}, and a limit theorem (e.g. Theorem~3 on page 351 in Konno~\cite{Konno2002a}) gives a limit distribution to the quantum walk.
The position of the walker divided by $4t$ converges in distribution.
For a real number $x$, we have
 \begin{align}
   &\lim_{t\to\infty}\mathbb{P}\left(\frac{X_t}{4t}\leq \frac{x}{4}\right)\nonumber\\
   =&\int_{-\infty}^{x/4} \frac{1}{\pi(1-4y^2)\sqrt{1-8y^2}}\nonumber\\
  & \qquad \times \biggl\{1-\Bigl(|\alpha|^2-|\beta|^2+2\alpha\overline{\beta}+2\overline{\alpha}\beta\Bigr)\cdot 2y\biggr\}I_{(-1/\sqrt{2}, 1/\sqrt{2})}(2y)\,d(2y)\nonumber\\
   =&\int_{-\infty}^{x} \frac{4\sqrt{2}}{\pi(4-y^2)\sqrt{2-y^2}}\nonumber\\
  & \qquad \times \biggl\{1-\Bigl(|\alpha|^2-|\beta|^2+2\alpha\overline{\beta}+2\overline{\alpha}\beta\Bigr)\frac{y}{2}\biggr\}I_{(-1/\sqrt{2}, 1/\sqrt{2})}\left(y/2\right)\,d\left(\frac{y}{2}\right)\nonumber\\
   =&\int_{-\infty}^{x} \frac{2\sqrt{2}}{\pi(4-y^2)\sqrt{2-y^2}}\biggl\{1-\Bigl(|\alpha|^2-|\beta|^2+2\alpha\overline{\beta}+2\overline{\alpha}\beta\Bigr)\frac{y}{2}\biggr\}I_{(-\sqrt{2}, \sqrt{2})}(y)\,dy,
 \end{align}
from which the statement of Theorem~\ref{th:limit_1} follows.

\subsection{$\theta\neq\pi/4, 3\pi/4$}
Differently from the case when the parameter $\theta$ is fixed at $\pi/4$ or $3\pi/4$, the one-step unitary operation $\hat{U}_2(k) \hat{U}_1(k)$ is not arranged in the same way as the previous discussion.
It is, however, possible to get a limit distribution for the quantum walk.

\begin{theorem}
Assume that $\theta\neq \pi/4, 3\pi/4$.
Let $c$ and $s$ be the short notations for $\cos\theta$ and $\sin\theta$, respectively.
For a real number $x$, we have
 \begin{align}
   &\lim_{t\to\infty}\mathbb{P}\left(\frac{X_t}{t}\leq x\right)\nonumber\\
   =&\int_{-\infty}^x \biggl\{f(y)\Bigl(1-\nu_{+}(\alpha,\beta; y)\Bigr)I_\mathcal{D}(y)+f(-y)\Bigl(1-\nu_{-}(\alpha,\beta; y)\Bigr)I_{\mathcal{D}}(-y)\biggr\}\,dy,
 \end{align}
 where
 \begin{align}
  f(x)=&\frac{\left(x+2\sqrt{D(x)}\right)^2}{2\pi(4-x^2)\sqrt{D(x)}\sqrt{W_{+}(x)}\sqrt{W_{-}(x)}},\\[3mm]
  D(x)=&1-4c^2s^2+c^2s^2x^2,\\
  W_{+}(x)=&-2(1-2|c|s)+(1-|c|s)x^2+x\sqrt{D(x)},\\
  W_{-}(x)=&2(1+2|c|s)-(1+|c|s)x^2-x\sqrt{D(x)},\\
 \end{align}
 \begin{align}
  \nu_{\pm}(\alpha,\beta; x)=&\frac{1}{2c^2-1}\biggl\{\left(c^2x\mp\sqrt{D(x)}\right)\left(|\alpha|^2-|\beta|^2\right)\nonumber\\
  &\qquad -\left(s^2x\mp\sqrt{D(x)}\right)\left(\alpha\overline{\beta}+\overline{\alpha}\beta\right)\biggr\},\label{eq:nu}\\
  \mathcal{D}=&\left(\sqrt{1-2|c|s},\,\sqrt{1+2|c|s}\right),\\
  I_{\mathcal{D}}(x)=&\left\{\begin{array}{cl}
	   1&(x\in \mathcal{D})\\
		  0&(x\notin \mathcal{D})
		 \end{array}\right..
 \end{align}
 \label{th:limit_2}
\end{theorem}

\begin{figure}[h]
\begin{center}
 \begin{minipage}{50mm}
  \begin{center}
   \includegraphics[scale=0.4]{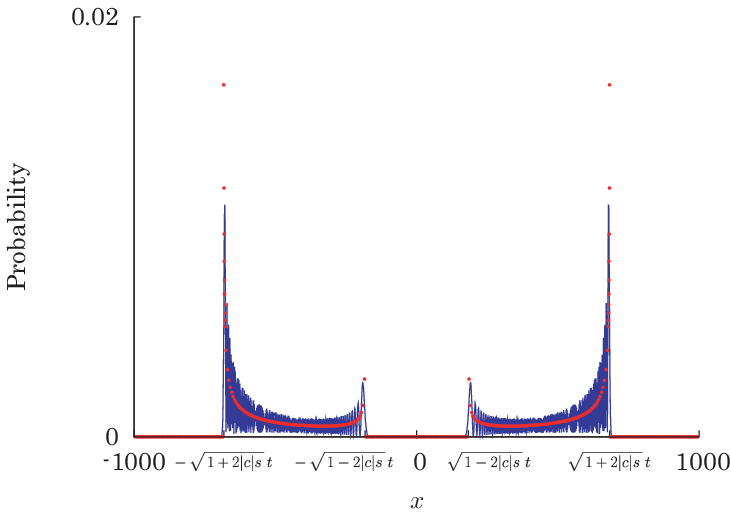}\\[2mm]
  (a) $\theta=\pi/6$
  \end{center}
 \end{minipage}
 \begin{minipage}{50mm}
  \begin{center}
   \includegraphics[scale=0.4]{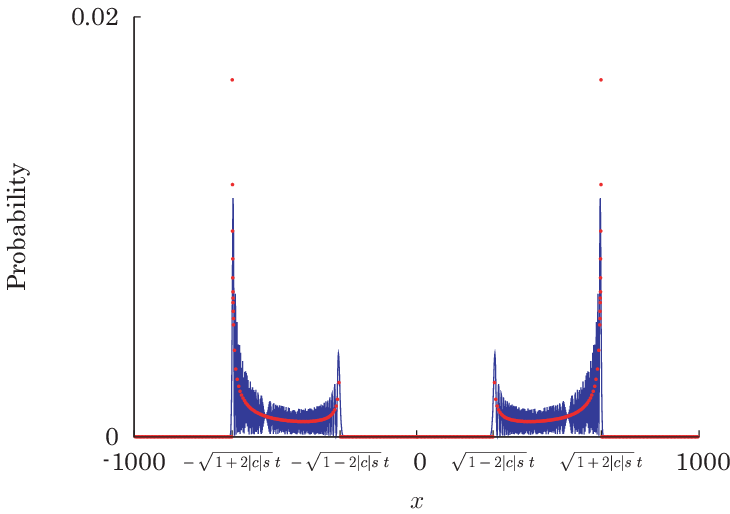}\\[2mm]
  (b) $\theta=\pi/8$
  \end{center}
 \end{minipage}
 \bigskip

\caption{(Color figure online) The blue lines represent the probability distribution $\mathbb{P}(X_t=x)$ at time $t=500$ and the red points represent the right side of Eq.~\eqref{eq:approximation-2} as $t=500$. The limit density function approximately reproduces the probability distribution as time $t$ becomes large enough. The walker launches off at the location $x=0$ as the initial state $\ket{\Psi_0}=\ket{0}\otimes (1/\sqrt{2}\ket{0}+i/\sqrt{2}\ket{1})$.}
\label{fig:5}
\end{center}
\end{figure}

The limit density function
\begin{equation}
 \chi_2(x)=f(x)\Bigl(1-\nu_{+}(\alpha,\beta; x)\Bigr)I_\mathcal{D}(x)+f(-x)\Bigl(1-\nu_{-}(\alpha,\beta; x)\Bigr)I_{\mathcal{D}}(-x),
\end{equation}
approximately estimates the probability distribution as time $t$ increases enough,
\begin{equation}
 \mathbb{P}(X_t=x)\sim \frac{1}{t}\cdot \chi_2\left(\frac{x}{t}\right)\quad (x\in\mathbb{Z})\label{eq:approximation-2}.
\end{equation}
The approximation nicely fits to the probability distribution, as shown in Fig.~\ref{fig:5}.

The proof of Theorem~\ref{th:limit_2} can be demonstrated by Fourier analysis.
We begin to decompose the initial state $\ket{\hat{\psi}_0(k)}=\alpha\ket{0}+\beta\ket{1}\,(=\ket{\phi})$ in the eigenspace of the unitary operation $\hat{U}_2(k)\hat{U}_1(k)$.
Let us represent the eigenvalues by $\lambda_j(k)\,(j=1,2)$ and the normalized eigenvectors by $\ket{v_j(k)}\,(j=1,2)$.
Note that the eigenvector corresponding to the eigenvalue $\lambda_j(k)$ should be denoted by $\ket{v_j(k)}$. 

The decomposition of the initial state
\begin{equation}
 \ket{\phi}=\sum_{j=1}^2 \braket{v_j(k)|\phi}\ket{v_j(k)},
\end{equation}
leads the Fourier transform of the quantum walk at time $t$ into the eigenspace,
\begin{equation}
 \ket{\hat\psi_t(k)}=\sum_{j=1}^2 \lambda_j(k)^t\braket{v_j(k)|\phi}\ket{v_j(k)},
\end{equation}
from which the $r$-th moments $\mathbb{E}[X_t^r]\, (r=0,1,2,\ldots)$ get forms in the eigenspace,
\begin{align}
 \mathbb{E}[X_t^r]
 =&\int_{-\pi}^{\pi} \bra{\hat\psi_t(k)} \Bigl(D^r\ket{\hat\psi_t(k)}\Bigr)\frac{dk}{2\pi}\nonumber\\
 =&(t)_r\left\{\sum_{j=1}^2 \int_{-\pi}^{\pi} \left(\frac{i\,\lambda'_j(k)}{\lambda_j(k)}\right)^r \Bigl|\braket{v_j(k)|\phi}\Bigr|^2 \frac{dk}{2\pi}\right\}+O(t^{r-1}),
\end{align}
where $D=i\cdot d/dk$ and $(t)_r=t\cdot (t-1)\times\cdots\times(t-r+1)$.
One can derive the limits of the $r$-th moments $\mathbb{E}[(X_t/t)^r]$ as $t\to\infty$,
\begin{equation}
 \lim_{t\to\infty}\mathbb{E}\left[\left(\frac{X_t}{t}\right)^r\right]
  =\sum_{j=1}^2 \int_{-\pi}^{\pi} \left(\frac{i\,\lambda'_j(k)}{\lambda_j(k)}\right)^r \Bigl|\braket{v_j(k)|\phi}\Bigr|^2 \frac{dk}{2\pi}.\label{eq:200807-3}
\end{equation}

Since the unitary operation $\hat{U}_2(k)\hat{U_1}(k)$ holds the eigenvalues
\begin{equation}
 \lambda_j(k)=\cos k + cs\sin^2 k-(-1)^j \sqrt{1-(\cos k + cs\sin^2 k)^2}\quad(j=1,2),
\end{equation}
we get
\begin{equation}
 \frac{i\,\lambda'_j(k)}{\lambda_j(k)}=(-1)^j\frac{\sin k}{|\sin k|}\cdot\frac{1-2cs\cos k}{\sqrt{(1-cs-cs\cos k)(1+cs-cs\cos k)}}\quad(j=1,2).
\end{equation}
With a function
\begin{equation}
 g(k)=\cos k + cs\sin^2 k,
\end{equation}
we have a possible representation of the normalized eigenvectors
\begin{align}
 \ket{v_j(k)}
 =&\frac{1}{\sqrt{N_j(k)}}\biggl[\Bigl\{cs\sin^2 k + i\,c(c-s\cos k)\sin k\Bigr\}\ket{0}\nonumber\\
 &+i\,\Bigl\{-(-1)^j\sqrt{1-g(k)^2} - s(s-c\cos k)\sin k\Bigr\}\ket{1}\biggr]\quad (j=1,2),
\end{align}
in which the notations $N_j(k)\,(j=1,2)$ denote the normalizing factors
\begin{equation}
 N_j(k)
 = c^2(1-2cs\cos k)\sin^2 k + \left\{-(-1)^j\sqrt{1-g(k)^2} - s(s-c\cos k)\sin k\right\}^2.
\end{equation}

Now, we focus on the $r$-th moment $\mathbb{E}[(X_t/t)^r]$ in Eq.~\eqref{eq:200807-3} and are going to compute it specifically so that the limit density function appears.
Introducing a function
\begin{equation}
 h(k)=\frac{1-2cs\cos k}{\sqrt{(1-cs-cs\cos k)(1+cs-cs\cos k)}},
\end{equation}
we multiple the right side of Eq.~\eqref{eq:200807-3} by $2\pi$ and split each integral into two parts,
\begin{align}
 & \int_{-\pi}^{\pi} \left(\frac{i\,\lambda'_1(k)}{\lambda_1(k)}\right)^r \Bigl|\braket{v_1(k)|\phi}\Bigr|^2\,dk\nonumber\\
 =& \int_{-\pi}^{0} \left(\frac{i\,\lambda'_1(k)}{\lambda_1(k)}\right)^r \Bigl|\braket{v_1(k)|\phi}\Bigr|^2\,dk + \int_{0}^{\pi} \left(\frac{i\,\lambda'_1(k)}{\lambda_1(k)}\right)^r \Bigl|\braket{v_1(k)|\phi}\Bigr|^2\,dk\nonumber\\
 =& \int_{-\pi}^{0} h(k)^r \Bigl|\braket{v_1(k)|\phi}\Bigr|^2\,dk + \int_{0}^{\pi} (-h(k))^r \Bigl|\braket{v_1(k)|\phi}\Bigr|^2\,dk\nonumber\\
 =& \int_{0}^{\pi} h(k)^r \Bigl|\braket{v_1(-k)|\phi}\Bigr|^2\,dk + \int_{0}^{\pi} (-h(k))^r \Bigl|\braket{v_1(k)|\phi}\Bigr|^2\,dk,\\[3mm]
 & \int_{-\pi}^{\pi} \left(\frac{i\,\lambda'_2(k)}{\lambda_2(k)}\right)^r \Bigl|\braket{v_2(k)|\phi}\Bigr|^2\,dk\nonumber\\
 =& \int_{-\pi}^{0} \left(\frac{i\,\lambda'_2(k)}{\lambda_2(k)}\right)^r \Bigl|\braket{v_2(k)|\phi}\Bigr|^2\,dk + \int_{0}^{\pi}\left(\frac{i\,\lambda'_2(k)}{\lambda_2(k)}\right)^r \Bigl|\braket{v_2(k)|\phi}\Bigr|^2\,dk\nonumber\\
 =& \int_{-\pi}^{0} (-h(k))^r \Bigl|\braket{v_2(k)|\phi}\Bigr|^2\,dk + \int_{0}^{\pi} h(k)^r \Bigl|\braket{v_2(k)|\phi}\Bigr|^2\,dk\nonumber\\
 =& \int_{0}^{\pi} (-h(k))^r \Bigl|\braket{v_2(-k)|\phi}\Bigr|^2\,dk + \int_{0}^{\pi} h(k)^r \Bigl|\braket{v_2(k)|\phi}\Bigr|^2\,dk.
\end{align}
These decompositions lead to integrals over the interval $(0, \pi)$ where the function $h(k)$ is strictly monotone,
\begin{align}
 & \sum_{j=1}^2 \int_{-\pi}^{\pi} \left(\frac{i\,\lambda'_j(k)}{\lambda_j(k)}\right)^r \Bigl|\braket{v_j(k)|\tilde\phi}\Bigr|^2\,dk\nonumber\\
 =& \int_{0}^{\pi} (-h(k))^r \left\{\Bigl|\braket{v_1(k)|\phi}\Bigr|^2 + \Bigl|\braket{v_2(-k)|\phi}\Bigr|^2\right\}\,dk\nonumber\\
& + \int_{0}^{\pi} h(k)^r \left\{\Bigl|\braket{v_2(k)|\phi}\Bigr|^2 + \Bigl|\braket{v_1(-k)|\phi}\Bigr|^2\right\}\,dk\nonumber\\
=& \int_{0}^{\pi} (-h(k))^r \left\{\Bigl|\braket{v_1(k)|\phi}\Bigr|^2 + \Bigl|\braket{v_2(-k)|\phi}\Bigr|^2\right\}\,\frac{1}{\frac{dh(k)}{dk}}\, dh(k)\nonumber\\
& + \int_{0}^{\pi} h(k)^r \left\{\Bigl|\braket{v_2(k)|\phi}\Bigr|^2 + \Bigl|\braket{v_1(-k)|\phi}\Bigr|^2\right\}\,\frac{1}{\frac{dh(k)}{dk}}\,dh(k).
\end{align}
With three functions
\begin{align}
 D(x)=&1-4c^2s^2+c^2s^2x^2,\\
 W_{+}(x)=&-2(1-2|c|s)+(1-|c|s)x^2+x\sqrt{D(x)},\\
 W_{-}(x)=&2(1+2|c|s)-(1+|c|s)x^2-x\sqrt{D(x)},
\end{align}
we have specific representations for the integrands,
\begin{align}
 & \Bigl|\braket{v_1(k)|\phi}\Bigr|^2 + \Bigl|\braket{v_2(-k)|\phi}\Bigr|^2\nonumber\\
 =& \left(1+\frac{c^2 h(k)-\sqrt{D(h(k))}}{2c^2-1}\right) |\alpha|^2
 + \left(1-\frac{c^2 h(k)-\sqrt{D(h(k))}}{2c^2-1}\right) |\beta|^2\nonumber\\
 & \qquad - \frac{s^2 h(k)-\sqrt{D(h(k))}}{2c^2-1} (\alpha\overline{\beta}+\overline{\alpha}\beta),\\
 & \Bigl|\braket{v_2(k)|\phi}\Bigr|^2 + \Bigl|\braket{v_1(-k)|\phi}\Bigr|^2\nonumber\\
 =& \left(1-\frac{c^2 h(k)-\sqrt{D(h(k))}}{2c^2-1}\right) |\alpha|^2
 + \left(1+\frac{c^2 h(k)-\sqrt{D(h(k))}}{2c^2-1}\right) |\beta|^2\nonumber\\
 & \qquad + \frac{s^2 h(k)-\sqrt{D(h(k))}}{2c^2-1} (\alpha\overline{\beta}+\overline{\alpha}\beta),
\end{align}
and
\begin{equation}
 \frac{dh(k)}{dk}= \frac{c (4-h(k)^2)\sqrt{D(h(k))}\sqrt{W_{+}(h(k))}\sqrt{W_{-}(h(k))}}{|c|\left(h(k)+2\sqrt{D(h(k))}\right)^2}.
\end{equation}

Noting $\lim_{k\to 0} h(k)=\sqrt{1-2cs}$ and $\lim_{k\to\pi} h(k)=\sqrt{1+2cs}$, we substitute $h(k)=x$.
Such a substitution brings the limit of the $r$-th moment $\mathbb{E}[(X_t/t)^r]$ to another integral form,
\begin{align}
 & \lim_{t\to\infty}\mathbb{E}\left[\left(\frac{X_t}{t}\right)^r\right]\nonumber\\[3mm]
 = & \int_{\sqrt{1-2cs}}^{\sqrt{1+2cs}} (-x)^r\, \Biggl\{\left(1+\frac{c^2 x-\sqrt{D(x)}}{2c^2-1}\right) |\alpha|^2\nonumber\\
 & \qquad + \left(1-\frac{c^2 x-\sqrt{D(x)}}{2c^2-1}\right) |\beta|^2  - \frac{s^2 x-\sqrt{D(x)}}{2c^2-1} (\alpha\overline{\beta}+\overline{\alpha}\beta)\Biggr\}\nonumber\\
 &\qquad \times \frac{|c|\left(x+2\sqrt{D(x)}\right)^2}{c (4-x^2)\sqrt{D(x)}\sqrt{W_{+}(x)}\sqrt{W_{-}(x)}}\cdot\frac{dx}{2\pi}\nonumber\\[3mm]
 & + \int_{\sqrt{1-2cs}}^{\sqrt{1+2cs}} x^r\, \Biggl\{\left(1-\frac{c^2 x-\sqrt{D(x)}}{2c^2-1}\right) |\alpha|^2\nonumber\\
 & \qquad + \left(1+\frac{c^2 x-\sqrt{D(x)}}{2c^2-1}\right) |\beta|^2  + \frac{s^2 x-\sqrt{D(x)}}{2c^2-1} (\alpha\overline{\beta}+\overline{\alpha}\beta)\Biggr\}\nonumber\\
 & \qquad \times \frac{|c|\left(x+2\sqrt{D(x)}\right)^2}{c (4-x^2)\sqrt{D(x)}\sqrt{W_{+}(x)}\sqrt{W_{-}(x)}}\cdot\frac{dx}{2\pi}\nonumber\\[3mm]
 = & \int_{-\sqrt{1+2cs}}^{-\sqrt{1-2cs}} x^r\,\Biggl[|\alpha|^2+|\beta|^2 -\frac{1}{2c^2-1}\Biggl\{\left(c^2x+\sqrt{D(x)}\right) (|\alpha|^2-|\beta|^2)\nonumber\\
 & \qquad - \left(s^2x+\sqrt{D(x)}\right) (\alpha\overline{\beta}+\overline{\alpha}\beta)\Biggr\}\frac{|c|}{c}f(-x)\,dx\nonumber\\
 & + \int_{\sqrt{1-2cs}}^{\sqrt{1+2cs}} x^r\,\,\Biggl[|\alpha|^2+|\beta|^2 -\frac{1}{2c^2-1}\Biggl\{\left(c^2x-\sqrt{D(x)}\right) (|\alpha|^2-|\beta|^2)\nonumber\\
 & \qquad - \left(s^2x-\sqrt{D(x)}\right) (\alpha\overline{\beta}+\overline{\alpha}\beta)\Biggr\}\frac{|c|}{c}f(x)\,dx\nonumber\\[3mm]
 = & \int_{-\sqrt{1+2|c|s}}^{-\sqrt{1-2|c|s}} x^r\,\Biggl[|\alpha|^2+|\beta|^2 -\frac{1}{2c^2-1}\Biggl\{\left(c^2x+\sqrt{D(x)}\right) (|\alpha|^2-|\beta|^2)\nonumber\\
 & \qquad - \left(s^2x+\sqrt{D(x)}\right) (\alpha\overline{\beta}+\overline{\alpha}\beta)\Biggr\}f(-x)\,dx\nonumber\\
 & + \int_{\sqrt{1-2|c|s}}^{\sqrt{1+2|c|s}} x^r\,\,\Biggl[|\alpha|^2+|\beta|^2 -\frac{1}{2c^2-1}\Biggl\{\left(c^2x-\sqrt{D(x)}\right) (|\alpha|^2-|\beta|^2)\nonumber\\
 & \qquad - \left(s^2x-\sqrt{D(x)}\right) (\alpha\overline{\beta}+\overline{\alpha}\beta)\Biggr\}f(x)\,dx,
\end{align}
where
\begin{equation}
 f(x)=\frac{\left(x+2\sqrt{D(x)}\right)^2}{2\pi(4-x^2)\sqrt{D(x)}\sqrt{W_{+}(x)}\sqrt{W_{-}(x)}}.
\end{equation}

Reminding the constraint $|\alpha|^2+|\beta|^2=1$, we find
\begin{align}
 & \lim_{t\to\infty}\mathbb{E}\left[\left(\frac{X_t}{t}\right)^r\right]\nonumber\\[3mm]
 = & \int_{-\sqrt{1+2|c|s}}^{-\sqrt{1-2|c|s}} x^r\,\Biggl[1 -\frac{1}{2c^2-1}\Biggl\{\left(c^2x+\sqrt{D(x)}\right) (|\alpha|^2-|\beta|^2)\nonumber\\
 & \qquad - \left(s^2x+\sqrt{D(x)}\right) (\alpha\overline{\beta}+\overline{\alpha}\beta)\Biggr\}f(-x)\,dx\nonumber\\
 & + \int_{\sqrt{1-2|c|s}}^{\sqrt{1+2|c|s}} x^r\,\Biggl[1 -\frac{1}{2c^2-1}\Biggl\{\left(c^2x-\sqrt{D(x)}\right) (|\alpha|^2-|\beta|^2)\nonumber\\
 & \qquad - \left(s^2x-\sqrt{D(x)}\right) (\alpha\overline{\beta}+\overline{\alpha}\beta)\Biggr\}f(x)\,dx,
\end{align}
and finally result in the desirable form,
\begin{align}
 & \lim_{t\to\infty}\mathbb{E}\left[\left(\frac{X_t}{t}\right)^r\right]\nonumber\\
 = & \int_{-\sqrt{1+2|c|s}}^{-\sqrt{1-2|c|s}} x^r\,f(-x)\Bigl(1-\nu_{-}(\alpha,\beta; x)\Bigr)\,dx\nonumber\\
 & + \int_{\sqrt{1-2|c|s}}^{\sqrt{1+2|c|s}} x^r\,f(x)\Bigl(1-\nu_{+}(\alpha,\beta; x)\Bigr)\,dx\nonumber\\
 = & \int_{-\infty}^{\infty} x^r\,\biggl\{f(-x)\Bigl(1-\nu_{-}(\alpha,\beta; x)\Bigr)I_{\mathcal{D}}(-x) + f(x)\Bigl(1-\nu_{+}(\alpha,\beta; x)\Bigr)I_\mathcal{D}(x)\biggr\}\,dx,
\end{align}
with
\begin{align}
 \nu_{\pm}(\alpha,\beta; x)=&\frac{1}{2c^2-1}\biggl\{\left(c^2x\mp\sqrt{D(x)}\right)\left(|\alpha|^2-|\beta|^2\right)\nonumber\\
 &\qquad -\left(s^2x\mp\sqrt{D(x)}\right)\left(\alpha\overline{\beta}+\overline{\alpha}\beta\right)\biggr\},\\
 \mathcal{D}=&\left(\sqrt{1-2|c|s},\,\sqrt{1+2|c|s}\right),\\
 I_{\mathcal{D}}(x)=&\left\{\begin{array}{cl}
		      1&(x\in \mathcal{D})\\
			     0&(x\notin \mathcal{D})
			    \end{array}\right..
\end{align}
The convergence of the $r$-th moments ($r=0,1,2,\ldots$) guarantees Theorem~\ref{th:limit_2}.

\section{Summary}
We studied a 1-dimensional quantum walk and discovered that the finding probability of the walker could hold a gap in distribution.
The system of quantum walk was operated by two unitary operations at each step.
The operations were characterized by a parameter $\theta\in (0,\pi)\,(\theta\neq\pi/2)$ and some numerical experiments showed that the walker seemed to be distributing with a gap around the launching location, except for the case $\theta=\pi/4, 3\pi/4$.
To prove the existence of the gap analytically, we derived long-time limit theorems (Theorems~\ref{th:limit_1} and \ref{th:limit_2}) and understood that the conjecture coming from the numerical calculation was correct.
Each of limit theorems stated a long-time limit distribution and precisely told us how the quantum walker distributed after repeating its update a lot of times.
Indeed, using the limit density function, we made an approximation to the probability distribution and it asymptotically reproduced the probability distribution.

The components of the unitary operations $U_1, U_2$ were under real numbers $c=\cos\theta, s=\sin\theta$ and did not have any phase terms.
We have not succeeded in computing a limit distribution for the quantum walk if the unitary operations include phase terms, due to the complexity of analysis.
It would be a future study to try to get a long-time limit theorem for such a quantum walk.

\backmatter

\bmhead{Acknowledgements}

The author is supported by JSPS Grant-in-Aid for Scientific Research (C) (No.23K03220).

\section*{Declarations}

Not applicable


\end{document}